\documentclass[12pt]{article}
\usepackage{epsf}
\usepackage{graphicx}
\usepackage{a4}
\usepackage{amsmath}
\usepackage{amssymb}
\usepackage{cite}		
\usepackage{color}
\usepackage{tikz}
\usepackage{pgfplots}
\usepackage{multicol}
\usepackage{multirow}
\usepackage{array}
\newcolumntype{P}[1]{>{\centering\arraybackslash}p{#1}}
\newcolumntype{M}[1]{>{\centering\arraybackslash}m{#1}}
\usepackage{empheq} 
\usepgfplotslibrary{fillbetween}

\usepackage{chemfig}

\catcode`\_=11
\definearrow5{-x>}{%
    \CF_arrowshiftnodes{#3}%
    \CF_expafter{\draw[}\CF_arrowcurrentstyle](\CF_arrowstartnode)--(\CF_arrowendnode)%
        coordinate[midway,shift=(\CF_ifempty{#4}{225}{#4}+\CF_arrowcurrentangle:\CF_ifempty{#5}{5pt}{#5})](line@start)%
        coordinate[midway,shift=(\CF_ifempty{#4}{45}{#4}+\CF_arrowcurrentangle:\CF_ifempty{#5}{5pt}{#5})](line@end)%
        coordinate[midway,shift=(\CF_ifempty{#4}{135}{#4}+\CF_arrowcurrentangle:\CF_ifempty{#5}{5pt}{#5})](line@start@i)%
        coordinate[midway,shift=(\CF_ifempty{#4}{315}{#4}+\CF_arrowcurrentangle:\CF_ifempty{#5}{5pt}{#5})](line@end@i);
    \draw(line@start)--(line@end);%
    \draw(line@start@i)--(line@end@i);%
    \CF_arrowdisplaylabel{#1}{0.5}+\CF_arrowstartnode{#2}{0.5}-\CF_arrowendnode
}
\catcode`\_=8
\def\hybrid{\topmargin 0pt
        \oddsidemargin 0pt
        \headheight 0pt \headsep 0pt
        \textwidth 6.25in       
        \textheight 9.5in       
        \marginparwidth .875in
        \parskip 5pt plus 1pt   \jot = 1.5ex}

\catcode`\@=11
\def\marginnote#1{}

\newcount\hour
\newcount\minute
\newtoks\amorpm
\hour=\time\divide\hour by60
\minute=\time{\multiply\hour by60 \global\advance\minute by-\hour}
\edef\standardtime{{\ifnum\hour<12 \global\amorpm={am}%
        \else\global\amorpm={pm}\advance\hour by-12 \fi
        \ifnum\hour=0 \hour=12 \fi
        \number\hour:\ifnum\minute<10 0\fi\number\minute\the\amorpm}}
\edef\militarytime{\number\hour:\ifnum\minute<10 0\fi\number\minute}

\def\draftlabel#1{{\@bsphack\if@filesw {\let\thepage\relax
   \xdef\@gtempa{\write\@auxout{\string
      \newlabel{#1}{{\@currentlabel}{\thepage}}}}}\@gtempa
   \if@nobreak \ifvmode\nobreak\fi\fi\fi\@esphack}
        \gdef\@eqnlabel{#1}}
\def\@eqnlabel{}
\def\@vacuum{}
\def\draftmarginnote#1{\marginpar{\raggedright\scriptsize\tt#1}}

\def\draft{\oddsidemargin -.5truein
        \def\@oddfoot{\sl preliminary draft \hfil
        \rm\thepage\hfil\sl\today\quad\militarytime}
        \let\@evenfoot\@oddfoot \overfullrule 3pt
        \let\label=\draftlabel
        \let\marginnote=\draftmarginnote
   \def\@eqnnum{(\theequation)\rlap{\kern\marginparsep\tt\@eqnlabel}%
\global\let\@eqnlabel\@vacuum}  }

\def\numberbysection{\@addtoreset{equation}{section}
        \def\theequation{\thesection.\arabic{equation}}}

\def\titlepage{\@restonecolfalse\if@twocolumn\@restonecoltrue\onecolumn
     \else \newpage \fi \thispagestyle{empty}\c@page\z@
        \def\thefootnote{\fnsymbol{footnote}}
	\setcounter{page}{0} }
\def\endtitlepage{\if@restonecol\twocolumn \else  \fi
        \def\thefootnote{\arabic{footnote}}
        \setcounter{footnote}{0}}  
\definecolor{c1}{rgb}{1, 0, 0}
\definecolor{c2}{rgb}{0, 1, 0}
\definecolor{c3}{rgb}{0, 0, 1}
\definecolor{c4}{rgb}{1, 0, 1}
\definecolor{c5}{rgb}{0, 1, 1}

\catcode`@=12
\relax

\def\ie{\hbox{\it i.e.}}
\def\nn{\nonumber}
\def \elr{\epsilon_{LR}}
\def \esr{\epsilon_{SR}}
\def \beq{\begin{equation}}
\def \eeq{\end{equation}}
\def\bea{\begin{eqnarray}}
\def\eea{\end{eqnarray}}
\def\EQ{\begin{equation}}
\def\EN{\end{equation}}

\relax
\numberbysection
\hybrid

\begin{document}

\begin{center}
{\large\bf Magnetic exponent for the long-range bond disordered Potts model}\\[.3in] 
{\bf Ivan\ Lecce \& Marco\ Picco}\\
  Sorbonne Universit\'e \& CNRS, UMR 7589, LPTHE, F-75005, Paris, France\\
    e-mail: {\tt lecce,picco@lpthe.jussieu.fr} \\
    {\bf Raoul\ Santachiara}\\
  Paris-Saclay Universit\'e \& CNRS, UMR 8626, LPTMS, 91405, Saclay, France\\
    e-mail: {\tt raoul.santachiara@gmail.com}
\end{center}
\centerline{(Dated: \today)}
\vskip .2in
\centerline{\bf ABSTRACT}
\begin{quotation}
We consider the critical behavior of two-dimensional Potts models in presence of a bond disorder
in which the correlation decays as a power law. In some recent work the thermal sector of this theory was investigated by a renormalization group computation based on perturbed conformal field theory. Here we apply the same approach to study instead the magnetic sector. In particular we compute the leading corrections to the Potts spin scaling dimension. Our results include as a special case the long-range disorder Ising model. We compare our prediction to Monte-Carlo simulations. Finally, by studying the magnetization scaling function, we show a clear numerical evidence of a cross-over between the long-range and the short-range class of universality.

\end{quotation}
\section{Introduction}
We consider here a two-dimensional $q-$Potts model on a square lattice in which the values of the couplings between spins are random. The disorder distribution, \ie\, the distribution from which the couplings are drawn,  exhibits long-range behavior, meaning that the correlation between distant couplings decreases according to a power law. We refer to this theory as the long-range disordered $q-$Potts model and denote the power law exponent by $a$. 

This model was investigated by Monte-Carlo methods in \cite{Chippari23} for $q=1,2,3$. It turned out that the critical behavior is governed, depending on the values of $q$ and $a$, by four different fixed points: the pure (P) point, where the disorder is irrelevant; the short-range (SR) point, where the short-range side of the disorder distribution dominates; and two points, the long-range (LR) and the infinite long-range  (ILR) points, to which the system is driven by the long-range side of the disorder distribution. At the LR point the disorder is finite, and the system is characterized by an interplay between thermal and disorder fluctuations while at the ILR point the thermal fluctuations are frozen. Similar observations were shown to be valid for the two-dimensional long-range disorder $q-$Potts on hierarchical lattices \cite{Wu_1994}, where a real-space renormalization group (RG) analysis support the existence of fixed points of type P, SR and LR. According to pertubative RG computations of Landau-Ginzburg actions, a similar scenario emerged for long-range disordered Potts model near their upper critical dimension $d_u=6$ \cite{Weinrib,
Stolan_84}  as well as for a family of long-range random mass multi-component $\phi^4$ models \cite{WeinribHalperin,Honkonen_1989}.

A perturbed conformal field theory approach to the two-dimensional long-range $q-$Potts model was introduced in \cite{Chippari23theo}. The thermal sector of the theory was investigated, in particular the central charge and the long-range correlation length exponent $\nu^{(LR)}$. The validity of the Weinrib-Halperin conjecture, according to which $ \nu^{(LR)}=2/a$, and the role played by the higher cumulants of the disorder distribution was clarified.    

In this paper we focus our attention to the magnetic sector. We derive in particular the exponent of the Potts spin pair correlation function.

\section{The lattice model}
We consider the Potts model on a square lattice with periodic boundary conditions on the two directions. The $q-$Potts model with bond disorder is defined by the partition function:
\begin{align}
\label{dpb}
&\mathcal{Z}(\{J_{<ij>}\})= \sum_{\{s_i\}}\; e^{-S\left(\{J_{<i,j>}\},\{s_i\}\right)}\\
&\text{with}\quad S\left(\{J_{<i,j>}\},\{s_i\}\right)=-\sum_{<ij>} J_{<ij>} \delta_{s_i, s_j}\; .
\end{align}
A spin $s_i$ is associated to the lattice vertex $i$ and it takes $q$ possible states, $s_i = \left\{ 1,\cdots, q \right\}$. The $<ij>$ identifies the edge connecting neighboring sites $i$ and $j$ and the $\delta_{k,l}$ is the Kronecker delta. Quenched bond disorder refers to the fact that the couplings $\{J_{<ij>}\}$ are random. Using a standard and, as explained below, convenient choice, we specialize to the situation where each coupling  $J_{<ij>}$ can take two positive values, $J_{<ij>}=J_1$ or $J_{<ij>}=J_2$, with equal probability. This is done by introducing an auxiliary random site variable $\sigma_i$, henceforth referred to as disorder variable,  which can take two values, $\sigma_i=-1,1$. We set  
\begin{align}
\label{bimodal}
J_{<ij>^{(R)}} =J_{<ij>^{(B)}}= \frac{J_1+J_2}{2} + \sigma_i \frac{J_1 - J_2}{2}\; ,
\end{align}
 where, the $J_{<ij>^{(R)}}$ and $J_{<ij>^{(B)}}$ are the couplings associated respectively to the edge on the right and to the edge on the bottom of a given site $i$. The pure (\ie\ no disorder) Potts model corresponds to the case when $J_1=J_2$. The Fortuin-Kasteleyn (FK) clusters can be constructed in a way strictly analogous to the one applied for pure Potts model. On each edge associated to the coupling $J_1$ ($J_2$) and connecting two equal spins, one put a bond with a probability $p_1=1-e^{-J_1}$ ($p_2=1-e^{-J_2}$). 

The bi-modal setup eq.~(\ref{bimodal}) is particularly convenient because the location of the critical point is exactly known \cite{kinzel1981critical}~:
\begin{equation}
\label{duality}
\left(e^{J_1}-1\right)\left(e^{J_2}-1\right)= q \; .
\end{equation}
So by fixing one coupling to satisfy eq.~(\ref{duality}), say $J_2=J_2(J_1,q)$,  we are left with one-parameter family of disordered critical models. We found convenient to use as parameter the following: 
\begin{equation}
\label{defmu}
\mu^2= \frac{\left(J_1-J_2(J_1,q)\right)^2}{4}, 
\end{equation}
because it is reminiscent of the strength disorder parameter used in the short range case. Notice that in \cite{Chippari23,Chippari23theo} we used instead the parameter $r=J_1/J_2$. The point $\mu^2=0$ corresponds to the pure (P) critical point. The $ \mu^2 = \infty$ is the infinite disordered point. In this limit we have, from eq.~(\ref{duality}), that  $J_2\to 0$ and $J_1\to\infty$. In this limit the  thermal fluctuations are frozen: the couplings configuration $\{J_{<ij>}\}$, or equivalently the set of the disorder variables $\{\sigma_{i}\}$, fixes the FK clusters configuration as $p_1=1-e^{-J_1}=1$, and $p_2=1-e^{-J_2}=0$. In particular the FK configuration coincides to the clusters of the edges with coupling $J_1$. 

The long-range nature of the disorder is related to the fact that the random variables are correlated over long distances. Concretely, the variables $\{\sigma_{i}\}$ are taken by some distribution such that:
 \begin{equation}
 \label{12cumulant}
 \mathbb{E}\left[\sigma_i\right]=0, \quad \mathbb{E}\left[\sigma_i \sigma_j\right]\sim  |i-j|^{-a} ,\quad |i-j|>>1
 \end{equation}  
where $\mathbb{E}\left[ \cdots\right]$ is the average over the $\{\sigma_i\}$ distribution.

Depending on the values of $q$ and $a$, the critical behavior of eq.~(\ref{dpb}) is described by the P point ($\mu^2=0$), by the SR point, ($\mu^2=\mu^2_{SR}$), by the LR fixed point (at finite disorder $\mu^2=\mu^2_{LR}$) and by the ILR point ($\mu^2=\infty$). In the latter case, depending on whether $a$ is greater or smaller than $3/2$, the ILR point is in the class of universality of the Bernoulli percolation or of the long-range percolation in which the critical exponents vary with $a$ \cite{Javerzat_2020}.  

An efficient way to generate long-range correlated random variables is to use the degrees of freedoms of auxiliary lattice models at criticality, see for instance \cite{Chatelain}. In the numerical simulations presented below, we generate the $\{\sigma_{i}\}$ configuration by using  $m-$copies of critical Ising model. 
The variable $\sigma_i$ is taken as a product of the spin variables $\sigma^{(\gamma)}$ of the $\gamma$ Ising copy :
\begin{equation}
\label{sigmamcopies}
\sigma_{i}=\prod_{\gamma=1}^{m} \sigma^{(\gamma)}_i
\end{equation}
Denoting $\mathbb{E}_{\gamma}\left[\cdots\right]$ the average under the critical $\gamma-$ Ising action, one as $\mathbb{E}_{\gamma}\left[\sigma^{(\gamma)}_i\right]=0$ and  $\mathbb{E}_{\gamma}\left[\sigma^{(\gamma)}_i\sigma^{(\gamma)}_j\right]\sim|i-j|^{-1/4}$. As $\mathbb{E}\left[\cdots\right]=\prod_{\gamma=1}^{m}  \mathbb{E}_{\gamma}\left[\cdots\right]$ (the Ising copies are not coupled), $\mathbb{E}\left[\sigma_{i}\right]=0$ and  $\mathbb{E}\left[\sigma_{i}\sigma_{j}\right]\sim|i-j|^{-m/4}$.  This fixes $a$ to be:
\begin{equation}
\label{amcopies}
a = \frac{m}{4}, \quad m=\text{Number of Ising models}.
\end{equation}
In our simulations  $a$ can vary along the above set of fractional numbers, $a=0.25,0.5,0.75\cdots$.
The disorder distribution generated in this way is not Gaussian. However, differently from the correlation length exponent $\nu^{(LR)}$ \cite{Chippari23theo}, we will show that the leading corrections to the magnetic exponent depend only on the first two cumulants. Our results, presented below, are therefore valid for all the distributions satisfying eq.~(\ref{12cumulant}).   
\section{The effective action at the continuum limit}
\label{subsec:effective_model_cont_limit}

The critical behavior of the system is captured by the continuum limit action:
\begin{equation}
\label{pertaction}
\mathcal{S}=\mathcal{S}^{\text{aux}}+\;\mathcal{S}^{\text{Potts}}+\;g^{0}_{LR} \int \;d^2 x\; \sigma(x) \varepsilon(x),
\end{equation} 
where $\mathcal{S}^{\text{Potts}}$ is conformal invariant action of the critical pure Potts model, $\sigma(x)$ and $\varepsilon(x)$ are respectively the disorder and the density energy field. They can be considered as the continuum limit of the lattice degree of freedoms $\sigma_{i}$ and $\delta_{s_i,s_j}$, $\sigma_{i}\to \sigma(x)$, $\delta_{s_i,s_j}\to \varepsilon(x)$. The $\mathcal{S}^{\text{aux}}$ fixes the probability distribution function of the $\sigma(x)$. In a perturbed conformal field theory approach, one does not need to specify the form of $\mathcal{S}^{\text{aux}}$, but it make assumptions on the behavior of its correlation function. For a long-range disorder, one can simply assume $\mathcal{S}^{\text{aux}}$ to be scale invariant such that the two-point correlation functions have an algebraic decay. This fixes the arena of theories to which we can apply our RG bootstrap approach. This approach is perfectly adapted to the disorder distribution used in our simulations where the action $\mathcal{S}^{\text{aux}}$ is the sum of $m-$ conformal Ising actions,  $\mathcal{S}^{\text{aux}}=\sum_{\gamma=1}^{m} \mathcal{S}^{\text{Ising}}$, see the discussion above eq.~(\ref{amcopies}).  

From eq.~(\ref{duality}) and comparing eq.~(\ref{defmu}) to eq.~(\ref{pertaction}), the coupling $g^{0}_{LR}$ is related to the disorder strength $\mu$ as:
\begin{equation}
\label{g0lr}
\left(g^{0}_{LR}\right)^2\propto \left(J_1-J_2\right)^2\propto \mu^2.
\end{equation}

To deal with quenched disorder, we consider the replicated action:
\begin{equation}
\label{replaction0}
\mathcal{S}^{(n)}=\mathcal{S}^{\text{aux}}+\;\mathcal{S}^{\text{($\alpha$)-Potts}}+\;g^{0}_{LR} \int \;d^2 x\; \sigma(x) \varepsilon^{(\alpha)}(x)
\end{equation}
In the above equation, the $\mathcal{S}^{\text{($\alpha$)-Potts}}$ is the action of the $\alpha-$ replica.  The theory eq.~(\ref{replaction0}) is not renormalizable as one needs another counterterm.  The renormalizable quantum field action is\cite{Chippari23theo}:   
 \begin{align}
\label{def:replicatedaction}
\mathcal{S}^{(n)}&=\mathcal{S}^{\text{aux}}+\sum_{\alpha=1}^{n}\;\mathcal{S}^{\text{($\alpha$)-Potts}}+ \mathcal{S}^{\text{pert}}\; , \\
 \mathcal{S}^{\text{pert}}&=\sum_{\alpha=1}^{n}\;g^{0}_{LR} \int \;d^2 x\; \sigma(x) \varepsilon^{(\alpha)}(x)+\sum_{\substack{\alpha,\beta=1 \\ \alpha\neq \beta}}^{n}\;g^{0}_{SR} \int \;d^2 x\; \varepsilon^{(\alpha)}(x) \varepsilon^{(\beta)}(x) \;.
 \end{align} 
 The part of the action $\mathcal{S}^{\text{pert}}$ which we consider as perturbation contains a long-range and a short-range term, associated respectively to the couplings $g^{0}_{LR}$ and $g^{0}_{SR}$. These couplings have dimensions $[g^{0}_{LR}]=2-\elr$ and $[g^{0}_{SR}]=2-\esr$ with:
\begin{equation}
\label{elresra}
\elr=1-\frac{a}{2}+\frac{\esr}{2}\; ,
\end{equation}
and 
\begin{equation}
\label{esrq}
\esr =4-\frac{6\pi}{2\pi-\arccos\left((q-2)/2\right)}\; =\frac{4}{3}\left(q-2\right)+O\left(\left(q-2\right)^2\right),
\end{equation}
see Appendix \ref{RGdetail}. 
The parameters $\elr$ and $\elr$ play the role of  RG  regularization parameters. The RG equations are valid in the regime where $\elr$ and $\elr$ are small, $\elr,\esr<<1$, or equivalently, $0<2-a<<1$ and $0<q-2<<1$. We denote as $g_{SR}$ and $g_{LR}$ the renormalized (dimensionless) coupling constants,  and $r$ the RG length scale. The RG flow of the couplings is described by the following equations\cite{Chippari23theo}:
\begin{align}
\label{beta_1loop}
\beta_{SR}&=r\frac{d}{dr} g_{SR}= \esr\; g_{SR} - 8\pi g_{SR}^2+\pi g_{LR}^2+O(g^3)\nonumber\\ 
 \beta_{LR}&= r\frac{d}{dr} g_{LR}=\elr\; g_{LR} - 4\pi g_{LR} g_{SR}+O(g^3),
\end{align}
where the symbol $O(x^3)$ is an abbreviation for $O(x^3)=O\left(x_{SR}^3, x_{LR}^3,x_{SR}^2x^{0}_{LR}\cdots\right)$. The RG flow (\ref{beta_1loop}) has three different fixed points $(g_{SR},g_{LR})=(g^{(X)}_{SR},g^{(X)}_{LR})$ where $X=P,SR,LR$:
\begin{align}
\label{PSR}
& g^{(P)}_{SR}=g^{(P)}_{LR}=0,\\
\label{SR}& g^{(SR)}_{SR}=\frac{\esr}{8\pi}+O(\esr^2),\; g^{(SR)}_{LR}=0 \\
\label{LR}
&g^{(LR)}_{SR}=\frac{\elr}{4\pi}+O(\elr^2,\elr\esr),\;g^{(LR)}_{LR}=\frac{1}{2\pi}\sqrt{\elr(2\elr-\esr)}+O(\epsilon^2).
\end{align}
 Notice that Eqs.~(\ref{beta_1loop})-(\ref{LR}) are the 1-loop results. The 2-loop results RG can be found in \cite{Chippari23theo} and depend on the fourth cumulant of the disorder distribution. Here, for computing the leading correction to the magnetic exponent, of order $O(\epsilon^3)$, we need to know the localization of the fixed points (\ref{SR}) and (\ref{LR}) at 1-loop order $O(\epsilon)$. This can be understood from the structure of the Potts spin renormalization bringing to eq.~(\ref{resgamma}). 
\section{Potts spin: RG results}
\label{PottspinRG}

The correlation length exponent $\nu^{(LR)}$ has been computed in \cite{Chippari23theo}. Here we derive the other critical exponent, the magnetic exponent $\beta^{(LR)}$. 
We use the scaling relation:
\begin{equation}
\label{betanuh}
\left(\frac{\beta}{\nu}\right)^{(X)}=h^{(X)}_{s}, 
\end{equation} 
where $h^{(X)}_{s}$ is the scaling dimension of the Potts spin at the critical point $X=P,SR,LR$, see eqs.~(\ref{PSR})-(\ref{LR}).  We compute the ratio $\left(\beta/\nu\right)^{(X)} $ by studying the renormalization of the Potts spin field. 

We note as $s^{(0)}_{\alpha}(x)$  the spin field of the $\alpha-$copy of the pure Potts model.  We compute the renormalization of the replica 
symmetric combination $\left(\sum_{\alpha=1}^{n} s^{(0)}_{\alpha}\right)$ and then take the limit $n\to 0$:
\begin{equation}
\label{Zs}
\left(\sum_{\alpha=1}^{n} s^{(\alpha)}\right) = Z_{s}\left(\sum_{\alpha=1}^{n} s_{(0)}^{(\alpha)}\right)
\end{equation}
where $Z_s= Z_{s}\left(r,n,g^{(0)}_{SR},g^{(0)}_{LR},\esr,\elr\right)$. 

From the solution of the Callan-Symanzyk equation at the RG fixed point, the scaling dimension $h^{\text{(X)}}_{s}$ at the fixed point eq.~(\ref{SR}), X$=SR$  and eq.~(\ref{LR}), X$=LR$, are given by~:
\begin{equation}
h^{\text{(X)}}_{s}=h^{(P)}_{s}-\gamma_{s}\left(g^{\text{(X)}}_{SR},g^{\text{(X)}}_{LR}\right),\quad \text{X}=SR,LR 
\end{equation}
where $h^{(P)}_s$ is the dimension at the pure point (\ref{PSR}) and:
\begin{equation}
\label{gammadef}
\gamma_s(g_{SR},g_{LR})=\lim_{n\to 0}\left[r\frac{1}{dr}\;\ln Z_{s}\left(r,n,g^{(0)}_{SR},g^{(0)}_{LR},\esr,\elr\right)\right].
\end{equation}
As usual, the fact that $\gamma_s$ can be expressed in terms of the renormalized coupling constants $g_{SR},g_{LR}$ is a manifestation of the renormalizability of the model. 
For the leading corrections to $h^{(LR)}_{s}$ we need to compute 3-loop diagrams. As shown in \cite{dotsenko1995renormalisation}, this was also the case for the $h^{(SR)}_{s}$.

We computed the $\gamma_{s}$ functions and we obtained:
\begin{align}
\label{resgamma}
\gamma_{s}\left(g_{SR},g_{LR})\right)&=  -\pi^2\;\left(\frac{2+\xi^2}{2} \right) \esr\;g_{SR}^2+ \frac{\pi^2}{4} \left(\elr+\frac{\xi^2}{2}\esr\right) \;g_{LR}^2 +\nonumber \\
&+ 8 \pi^3 \;g_{SR}^3-2 \frac{\elr}{2\elr-\esr}\;g_{SR}g_{LR}^2 +O(g^4) \; .
\end{align}
The $\xi$ is a constant given by:
\begin{equation}
\label{xi}
\xi=2\frac{\Gamma\left[\frac16\right] \Gamma\left[-\frac{2}{3}\right]}{\Gamma\left[-\frac16\right]\Gamma\left[-\frac13\right]}. 
\end{equation} 
The details for obtaining the result given in eq.~(\ref{resgamma}) are in Appendix \ref{RGdetail}.

At the SR point, using eq.~(\ref{betanuh}), one obtains:
\begin{equation}
\label{betanusr}
\left(\frac{\beta}{\nu}\right)^{SR}=\left(\frac{\beta}{\nu}\right)^{P}+\frac{1}{128} \esr^3 \xi^2,
\end{equation}
This is the result of \cite{dotsenko1995renormalisation}.

The value of the magnetic exponent at the LR is the main theoretical result in this work:
\begin{equation}
\label{betanuq}
\left(\frac{\beta}{\nu}\right)^{LR}=\left(\frac{\beta}{\nu}\right)^{P}-\frac{1}{32} \elr\;\left(\elr-\esr\right)\left(4\elr+\esr \;\xi^2\right)+O\left(\epsilon^4\right).
\end{equation}
Using eqs.~(\ref{elresra}) and (\ref{esrq}), the above result can be expressed in terms of the initial parameters of the problem, \ie\ the states $q$ of the Potts model and the exponent $a$ of the algebraic decay of the correlation between distant couplings. When applied to the Ising point (\ie\ $q=2$), we get:
\begin{equation}
\label{betanuising}
\text{ISING}:\quad\left(\frac{\beta}{\nu}\right)^{LR}=\frac{1}{8}-\frac{1}{64}\left(2-a\right)^3 +O\left((2-a)^4\right),\quad a\leq 2
\end{equation}

\begin{figure}[!ht]
\begin{center}
\includegraphics[angle=0,width=0.45\linewidth,clip=true]{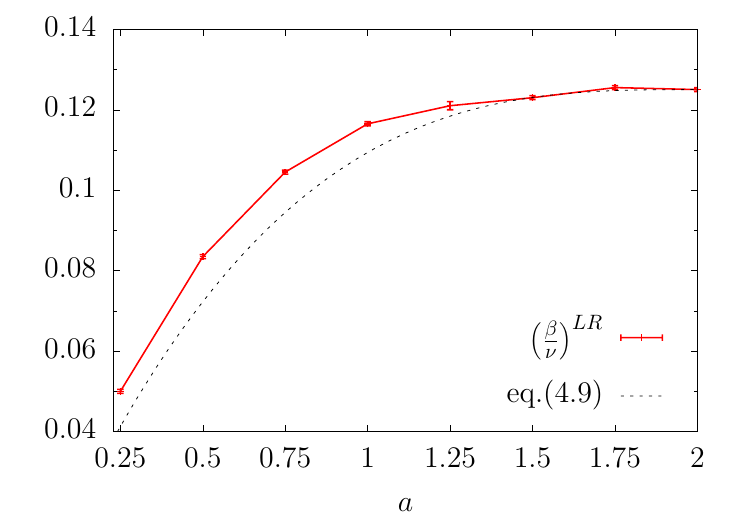}
\includegraphics[angle=0,width=0.45\linewidth,clip=true]{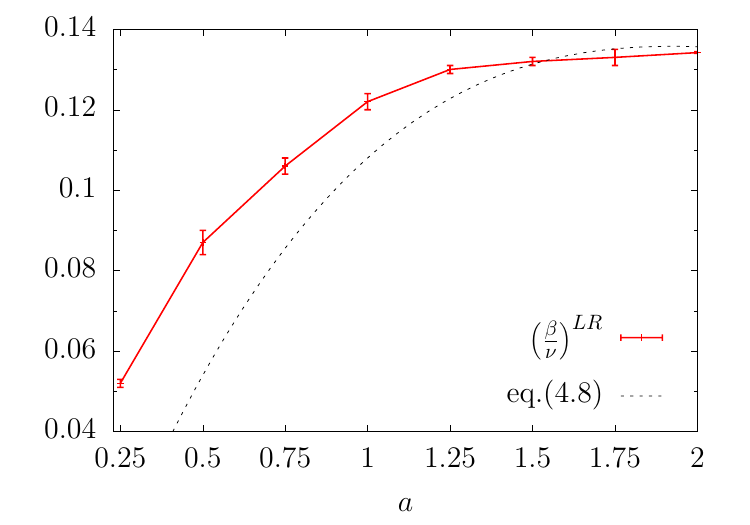}
\end{center}
\caption{Measured value $(\beta/\nu)^{LR}$ vs. $a$ for the $q=2$ on the left and for $q=3$ on the right.}
    \label{fig:etaq2}
\end{figure}

In the left part of Fig.~\ref{fig:etaq2} we compare the measured value $\left(\beta/\nu\right)^{LR}$ for $q=2$ in \cite{Chippari23}\footnote{Note that 
$(\beta/\nu)^{(LR)}$ corresponds to $\eta_1/2$, where  $\eta_1$ is the exponent appearing in Table 4 of this paper} with eq.~(\ref{betanuising}). 
The agreement is very good for $a\lesssim 2$ and deviates for smaller values of $a \leq 1$, as expected. A similar agreement between 
the theory and Monte-Carlo was observed in \cite{Chippari23}  for another exponent, the exponent of the square of the Ising spin correlation. 
For this exponent, the analytical predictions were given in \cite{dudka2016critical}. 

We also verified eq.~(\ref{betanuq}) by obtaining new Monte-Carlo results for $q=3$, reported in Table~\ref{q3nresult}. 
This table contains the values of $\left(\beta/\nu\right)^{LR}$ for different values of $a$ measured close to new critical fixed point and the values 
$\left(\beta/\nu\right)^{ILR}$ at infinite disorder. 
The errors for the LR are mostly due to the (un) determination of the value of disorder corresponding to the  fixed point (see a discussion on this point in next section).
For $a=2$, we show in fact the value obtained for a measurement with short range disorder since one expects that this is dominant over the long range disorder. 
For $0.75 \leq a \leq 1.75$, we always obtain 
a LR fixed point with a finite value of disorder and a magnetic exponent  larger than the one for the infinite disorder case thus confirming the previous 
results \cite{Chippari23} that the ILR point is unstable in this range. For $a \leq 0.5$, on the contrary, for any value of disorder, the magnetic exponent tends 
to the one of the ILR. The reported values in the table bellow, for $a=0.25$ and $0.5$, are for $\mu^2_{LR} = 4.4256$ and give the same exponents as for the infinite disorder. 
In particular, the measured value for $a=0.25$ is compatible with the exact value $\beta/\nu = 5/96$ determined previously \cite{Chippari23}. 
The agreement between theory and numerical results, shown in the right part of Fig.~\ref{fig:etaq2}, remains good for $a$ not too small.
\begin{table}[h]
\centering
\begin{tabular}{|P{2.cm}|P{2.9cm}|P{3.4cm}|}
\hline
$a$   & $\left(\beta/\nu\right)^{LR}$  & $\left(\beta/\nu\right)^{ILR}$ \\
\hline
 0.25  & $0.052 \pm 0.001$ &  $0.052 \pm 0.001$ \\
 \hline
 0.50  & $0.087 \pm 0.002$ &  $0.086 \pm 0.001$ \\
 \hline
 0.75  & $0.107 \pm 0.002$ & $0.102 \pm 0.001$  \\
 \hline
 1.00  & $0.122 \pm 0.002$ & $0.109 \pm 0.001$ \\
 \hline
1.25  &  $0.130 \pm 0.001$  & $0.107 \pm 0.001$ \\
\hline
1.50  & $0.132 \pm 0.001$  & $0.105 \pm 0.001$ \\
\hline
1.75  & $0.133 \pm 0.002 $  & $0.104 \pm 0.001$ \\
\hline
2.00  &  $0.1342 \pm 0.0002$  & $0.104 \pm 0.001$ \\
\hline
\end{tabular}
\caption{Magnetic exponent for the $q=3$ Potts model as a function of $a$ for the LR point and the ILR point.}
\label{q3nresult}
\end{table}
%
%
In the following section, we discuss some issues related to the measure of $(\beta/\nu)^{LR}$.
\section{Potts spin: Monte-Carlo measures }
\label{PottsspinMC}
The simplest method to measure the magnetic exponent is by using the relation between the value $\beta/\nu$ and the fractal dimension $D_{FK}$ of the FK cluster, $D_f=2-\beta/\nu$. 
As we sit at the percolation transition eq.~(\ref{duality}), the linear size of the largest FK cluster  scales as lattice size $L$, and its area $A$ scales, 
for each disorder configuration,  as $A\sim L^{D_f}\sim L^{2-\beta/\nu}$. We derive $\beta/\nu$ by measuring the following quantity:
\begin{equation}
m=\mathbb{E}\left[ \frac{A}{L^2}\right],
\end{equation}
where the $\mathbb{E}$ is the average over the set $\{\sigma_i\}$ and computing the effective exponent as
\begin{align}
\label{betanueff}
\frac{\beta}{\nu}\left(L\right) = - \frac{\log{\left(\frac{m(2L)}{m(L)}\right)}}{\log{(2)}}.
\end{align} 
At the percolation point eq.~(\ref{duality}), $m$ depends only on the size of the system and on the disorder strength $\mu^2$, see eq.~(\ref{defmu}), thus $m=m(\mu^2,L)$. 
As shown in a previous work, see \cite{Chippari23},  
depending on the value of $q$ and $a$, the system flows to one of the fixed points, $\mu^2=0$, $\mu^2=\mu_{X}^2$, $X=SR,LR$ or $\mu^2=\infty$. 

A precise determination of the value $\mu_{LR}^2$ is in fact difficult. The reason is that the magnetic exponent is affected by a 
dependency in the disorder $\mu^2$ and also by correction-to-scaling. This then leads to two types of corrections which can add or cancel each other,  
making the determination of the new critical exponent not very precise. 

\begin{figure}[!ht]
\begin{center}
\includegraphics[angle=0,width=0.8\linewidth,clip=true]{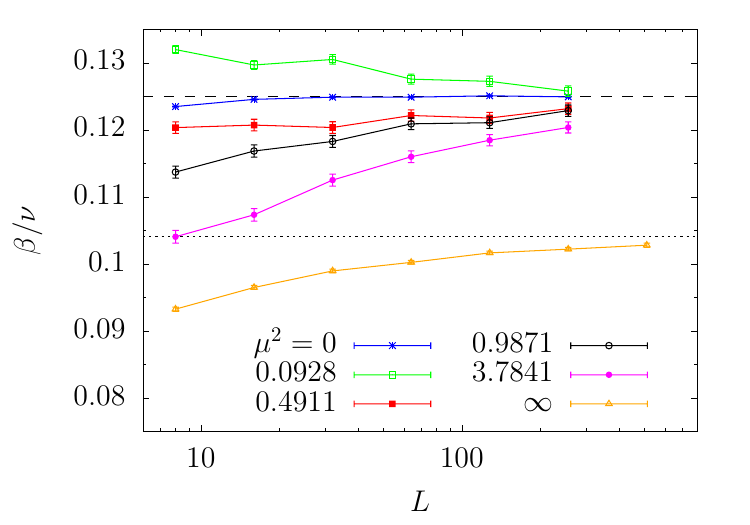}
\end{center}
\caption{Effective magnetic exponent vs. $L$ for the $q=2$ Potts model with $a=1.5$.}
    \label{fig:magq2a1.5}
\end{figure}

For the sake of clarity, let us discuss these issues for the Ising case, $q=2$ and for $a=1.5$. The corresponding numerical results are illustrated in Fig.~\ref{fig:magq2a1.5} where the effective exponent $\beta/\nu (L)$ is plotted as a function of $L$ for various values of disorder. When $\mu^2=0$ (pure Ising model), the effective exponent  converges to the Ising value $\left(\beta/\nu\right)^{P}=1/8$, $\lim_{L\to \infty}\beta/\nu (L)\to\left(\beta/\nu\right)^{P}$, shown as a dashed line, with a small deviation at small distances. 
It can be easily checked that this corresponds to corrections to the scaling of the form:
\begin{align}
\label{corsc}
\frac{\beta}{\nu}\left(L\right) = 0.125 + \alpha L^{-\omega} \; ,
\end{align}
with the correction-to-scaling exponent $\omega = 1.75$ and $\alpha$ some non-universal numerical factor. A fit of our data gives $\omega = 1.7 \pm 0.2$. For the case of the infinite disorder point ILR, $\mu^2 = \infty$, the effective exponent $\beta/\nu(L)$ 
converges to $\left(\beta/\nu\right)^{ILR}=5/48$,  $\lim_{L\to \infty}\beta/\nu (L)\to \left(\beta/\nu\right)^{ILR}$, shown as a dotted line. The value $5/48$ corresponds to the well known exponent for the uncorrelated percolation. This is also expected as, according to the extended Harris criterion \cite{Weinrib}, for the value $a=1.5$, the ILR main critical exponent coincides with the one of the  uncorrelated critical percolation. A similar fit to the form (\ref{corsc}) gives $\omega = 0.51 \pm 0.01$. Note that this correction-to-scaling exponent is much smaller than the one of the uncorrelated bond percolation for which a value $\simeq 3/2$ is known \cite{Ziff}. For the other values of disorder $\mu^2 \in[0.0928:3.784]$, the situation is less clear. In each case, 
$(\beta/\nu)(L)$ converge to a similar value $\simeq 0.12$. But the correction to the scaling strongly depends on the value of the disorder. 
For small disorder, \ie\ $\mu^2=0.0928$, the effective exponent decreases. Note that at small distances, the effective exponent is larger than for the pure model. Then for large sizes, it goes to a value $\simeq 0.125$.  For the strongest finite disorder considered, $\mu^2=3.784$, the effective exponent increases, starting from a value close to the one of infinite disorder, up to a value $\simeq 0.120$. 
Thus for all the finite disorder, we obtain for the largest size simulated (corresponding to $L=256$ and $512$),  $(\beta/\nu)^{LR}$ in the range $0.12 - 0.125$. The behavior just described can be explained  by an attractive fixed point for which one has 
\begin{align}
\label{corsc2}
\frac{\beta}{\nu}(\mu^2,L) &= g((\mu^2 - \mu_{LR}^2) L ^{y_d}) (1 + \alpha L^{-\omega}) \nn \\
                                         &=  g(0) + \rho (\mu^2 - \mu_{LR}^2) L ^{y_d} + \alpha L^{-\omega} + \cdots \; .
\end{align}
Since $y_d< 0$ (it is an attractive fixed point), one has two similar correction $(\mu^2 - \mu_{LR}^2) \rho L^{y_d}$ and $\alpha L^{-\omega}$ with the first one changing sign 
for $\mu = \mu_{LR}$.  For the smallest value of disorder, $\mu^2=0.0928$, the  $\beta/\nu(L)$ decreases. For the largest 
value of disorder, $\mu^2=3.784$, the  $\beta/\nu(L)$ increases. 
 
Since we do not know how to determine $\mu_{LR}^2$, $y_d$ or $\omega$, it is difficult to determine if the asymptotic value of $(\beta/\nu)^{LR}$ 
is $0.125$ or a smaller value, as expected if the long range disorder disorder is relevant. 
In \cite{Chippari23}, the value $(\beta/\nu)^{LR} = 0.123$ was determined for $a=1.5$ by using another type of measurement involving 
the computation of two points correlation function, allowing us to confirm that the disorder is relevant. 

The above observations motivated us in studying a rescaled version of the magnetisation, namely $\frac{m(\mu^2, L ) }{m(0, L )}$ for small values of the disorder $\mu^2$. The crucial point  
for studying this quantity is the assumption that the correction-to-scaling do not depend much on the disorder. By considering the rescaled magnetisation for small values of the disorder,
one expects that these corrections cancel. This allows to consider only the effect of the disorder close to the pure P point. 
By general scaling arguments we can then expect  that  
   \begin{equation}
      \frac{m(\mu^2, L ) }{m(0, L )}= f(\mu^2 L^{y_d}) \; .
  \end{equation}
The above scaling relation allows us to extract the renormalization exponent $y_d$ which is associated to the disorder perturbation. 
Due to the relation eq.~(\ref{g0lr}), we expect $y_d$, which is the RG eigenvalue associated to $\mu^2$, to be related to the scaling dimension of $(g^{0}_{LR})^2$ by:
\begin{equation}
\label{ydpred}
y_d = 2 \epsilon_{LR}.
\end{equation}
Below we present numerical results that are in very good agreement with this prediction. We discuss also the case $q=1$ to which the RG results presented in section~(\ref{PottspinRG}) cannot be applied. 

\subsection{Numerical results}
\label{numericalresults}
\subsubsection*{$q=1$} 
We start with the case $q=1$. On the lattice, all the spins take the same value. A bond  is present on the link $<ij>$ with a probability $1-e^{-J_{<ij>}}$. For each configuration of disorder $J_{<ij>}$, one construct the clusters by connecting these bonds, which gives a cluster configuration. This construction is very simple and fast to perform, so we can average 
over $N= 10^7$ samples of disorder. Simulations have been done for $L \in \{8, 16, 32, 64, 128, 256\}$ for all the values of $a$ in eq.~(\ref{amcopies}) with $m=1,\cdots, 8$.  
For the extended Harris criterion \cite{Weinrib}, and confirmed in \cite{Chippari23}, the disorder is relevant for $a \leq 1.5$, \ie\ $m \leq 6$. In each case, we found that there is a flow towards an ILR point $\mu^2=\infty$. 
Here we are interested in studying the start of the flow. Thus we consider weak disorder with $\mu^2 \in [0,0.2337]$. 

\begin{figure}[h]
 \includegraphics[angle=0,width=0.452\linewidth,clip=true]{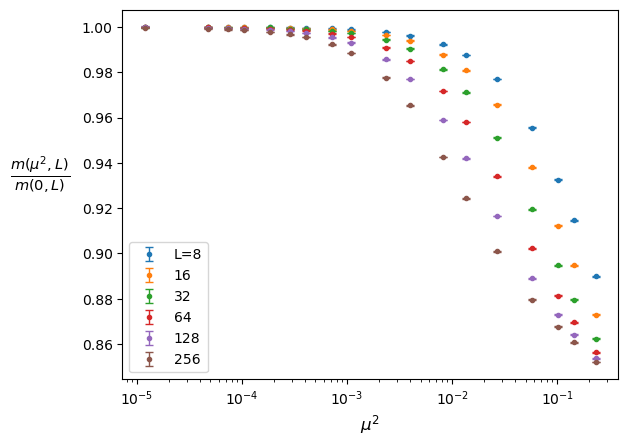}
 \includegraphics[angle=0,width=0.4\linewidth,clip=true]{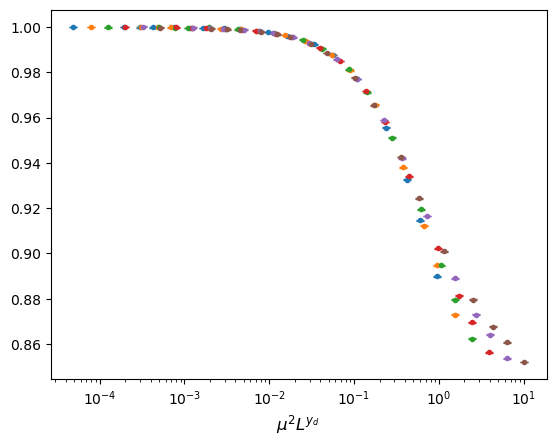}
    \caption{Left panel : Normalised magnetizations as a function of the disorder $\mu^2$ for $a=0.75$. Right panel : Same data as a function of $\mu^2 L^{y_d}$ with $y_d = 0.68$.}
    \label{fig:scaleQ1}
\end{figure}
In Fig.~\ref{fig:scaleQ1}, we show our results of the measurements of the normalised magnetisation $\frac{m(\mu^2, L ) }{m(0, L )}$ for $a=0.75$. In the left panel, we show it as a function of the disorder $\mu^2$. 
For a given amount of disorder, we clearly observe that the normalised magnetisation decreases as we increase the size $L$. This is in agreement with the claim that the disorder is relevant. 
The larger is the size, the more the data deviates from the pure case. Then, in the right panel, we check that, by rescaling the argument, the data collapse on a single curve. We obtain a nice collapse 
for $\mu^2 L^{y_d}$ up to $O(1)$ for $y_d = 0.68$. This value is obtained  through visual inspection. 
We checked that a similar collapse can be obtained for all values of $a$ with a different value of $y_d$. Before presenting these results for all $a$'s, let us explain in more details 
how we proceeded in order to obtain a more precise value for $y_d$, and compare it with scaling predictions. We consider the quantity 
\begin{align}
1-\frac{m(\mu^2,L)}{m(0,L)} 
\end{align} 
for two successive linear sizes $L$ and $2 L$, and plot it as a function of $\mu^2 L^{y_d}$. For each couple of pairs $L,2 L$, we adjust $y_d$ such as to minimize the distance between the two curves. 

\begin{figure}[h]
    \makebox[\textwidth][c]{\includegraphics[width=0.8\textwidth,  height=12cm]{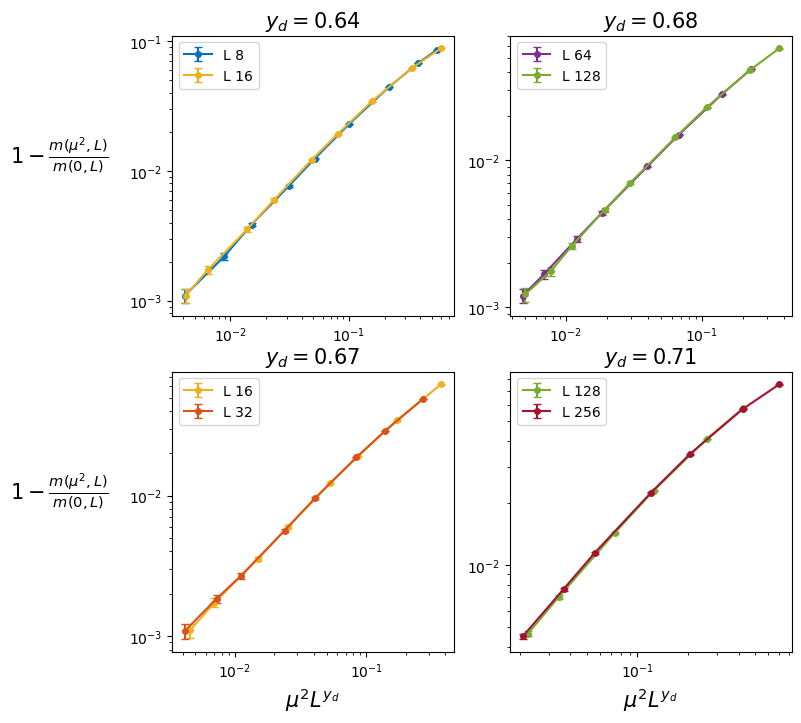}}%
    \caption{Collapse of the data $1-\frac{m(\mu^2,L)}{m(0,L)}$ vs. $\mu^2 L^{y_d}$ for the couples of curves with the lattice sizes as shown in the caption keys.}
    \label{fig:Q1Col}
\end{figure}

In Fig.~\ref{fig:Q1Col}, we show four different pairs of curves, with the corresponding numerical scaling exponent, in a double logarithmic scale. 
In this figure, one observes a variation of the critical exponent while increasing the lattice sizes. This can be explained in terms of additional finite size effects. 
This method has been applied for all the values of $a$ and compared with the theoretical predictions. Our results have been summed up in the Table~\ref{tableq1} where 
we show the results for the largest couple of lattice size, $L=128$ and $256$.

A numerical uncertainty to each result has been computed by changing the exponent up to a deviation of $20 \%$ with respect to the minimum distance between curves. 

Form the previous figure one can observe a variation of the critical exponent at increasing lattice size. This can be explained in terms of additional finite size effects. 
This method has been applied for all the values of $a$ and compared with the theoretical predictions. 
 
\begin{figure}[h]
    \makebox[\textwidth][c]{\includegraphics[width=11cm, height=9cm]{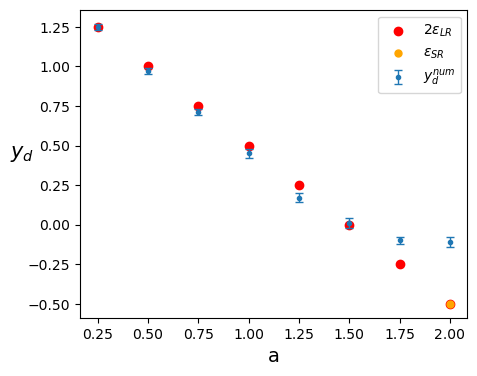}}%
    \caption{Comparison between the measure value $y_d^{num}$ and the predicted value for the long-range case, $2 \epsilon_{LR}= 3/2-a$, and for the short-range case, $\epsilon_{SR}=-0.5$.}
    \label{fig:ResQ1}
\end{figure}

\begin{table}[h]
\centering
\begin{tabular}{|c ||c |c|}
 \hline
 a & $y_d^{num}$   \\ [0.5ex] 
 \hline\hline
 0.25 & 1.25 $\pm$  0.02 \\ 
 \hline
 0.5 & 0.97 $\pm$ 0.02  \\
 \hline
 0.75 & 0.71 $\pm$ 0.02 \\
 \hline
 1 & 0.45 $\pm$ 0.03  \\
 \hline
  1.25 & 0.17 $\pm$ 0.03  \\ 
 \hline
 1.5 & 0.01$\pm$ 0.03  \\
 \hline
 1.75 &  -0.1 $\pm$ 0.02 \\
 \hline
 2 &  -0.12 $\pm$ 0.03 \\ 
 \hline
\end{tabular}
\caption{Numerical $y_d$ obtained for $L=128$ and $256$ for the $q=1$ Potts model.}
\label{tableq1} 
\end{table}
 We observe that the values of $y_d$ are close to the prediction~(\ref{ydpred}) for all values of $a \leq 1.5$, see Fig.~\ref{fig:ResQ1}. For $a > 1.5$, we obtain negative value for $y_d$, which is expected by the extended Harris criterion according to which the long-range disorder is irrelevant. In that case, the finite size corrections are much stronger.
For example, we measure for $a=2$, $y_d=-0.29$ for $L=8,16$, $-0.18$ for $L=32,64$ and $y_d=-0.12$ for $L=128,256$. An extrapolation is difficult to perform, but a result $y_d =0 $ is plausible. For higher values of $a$, it seems that $y_d$ goes to a zero value corresponding to a marginal case. 
\subsubsection*{q=2}
\begin{figure}[h]
    \makebox[\textwidth][c]{\includegraphics[width=11cm, height=9cm]{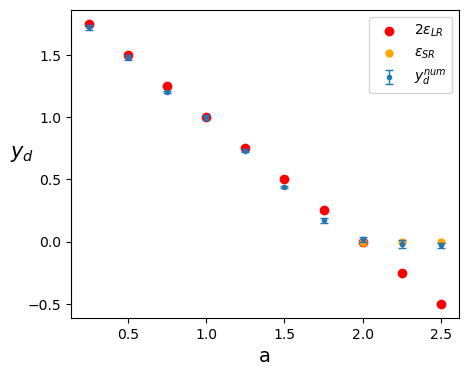}}%
    \caption{Comparison between the measure value $y_d^{num}$ and the predicted value for the long-range case, $2 \epsilon_{LR}=2-a$, and for the short-range case, $\epsilon_{SR}=0.0$.}
    \label{fig:ResQ2}
\end{figure}

For $q=2$, where one need to perform thermal averages over the spin configurations,  the number of samples has been reduced to $N=10^6$ while the range for the disorder strength $\mu^2 \in [0,0.2337]$ has been left unchanged.
The largest linear size considered is $L=128$ for a trade-off between the accuracy argument presented before and the increased running time of simulations. The result shown in Table~\ref{tableq2}  have been obtained by considering data for $ L \in \{64, 128\}$. Finally two additional values of 
$a=2.25, 2.5$ are considered with respect to ones udes in the $q=1$ simulations.
The  technique shown before has been applied for each correlation exponent and the final results for $y_d$ values are shown in the table below with the corresponding graphical representation in Fig.~\ref{fig:ResQ2}.

\begin{table}[h]
\centering
\begin{tabular}{|c || c |c||} 
 \hline
 a & $y_d$ \\ [0.5ex] 
 \hline\hline
 0.25 &  1.72 $\pm$ 0.02   \\ 
 \hline
 0.5 & 1.48 $\pm$ 0.02  \\
 \hline
 0.75 & 1.2 $\pm$ 0.01   \\
 \hline
 1 & 1 $\pm$  0.02  \\
 \hline
  1.25 & 0.73 $\pm$ 0.01   \\ 
 \hline
 1.5 & 0.44 $\pm$ 0.01   \\
 \hline
 1.75 & 0.17 $\pm  $ 0.02   \\
 \hline
 2 & 0.02 $\pm$ 0.02 \\
 \hline
  2.25 & -0.02  $\pm$ 0.03 \\
 \hline
 2.5 & -0.03  $\pm$ 0.02  \\ 
 \hline
\end{tabular}
\caption{Numerical $y_d$ obtained for $L=64$ and $128$ for the $q=2$ Potts model.}
\label{tableq2} 
\end{table}

The disorder critical exponent is relevant for $a<2$ and marginal for $a \ge 2$. When these values are compared with the theoretical prediction, the first important remark is 
related to the effective change of the disorder type captured by numerical results. For $a>2$ as expected, there is a change between long-range disorder to short-range one. 

\subsubsection*{q=3}
\begin{figure}[h]
    \makebox[\textwidth][c]{\includegraphics[width=11cm, height=9cm]{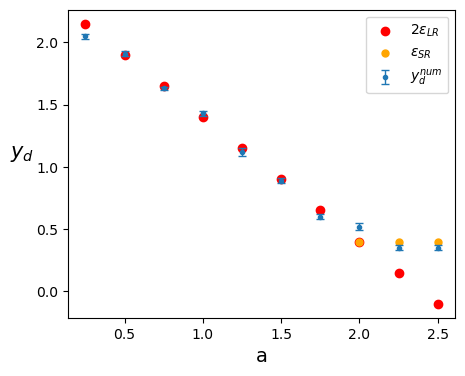}}%
    \caption{Comparison between the measure value $y_d^{num}$ and the predicted value for the long-range case, $2 \epsilon_{LR}=12/5-a$, and for the short-range case, $\epsilon_{SR}=0.4$.}
    \label{fig:ResQ3}
\end{figure}

Finally, we report our results for $q=3$, using the same parameters as for $q=2$ in table~\ref{tableq3}, again measured for $L=64-128$. We also show $y_d$ as a function of $a$ in Fig.~\ref{fig:ResQ3}. 
\begin{table}[h]
\centering
\begin{tabular}{|c || c |} 
 \hline
 a & $y_d$   \\ [0.5ex] 
 \hline\hline
 0.25 &  2.05 $\pm$ 0.02  \\ 
 \hline
 0.5 & 1.91 $\pm$ 0.02   \\
 \hline
 0.75 & 1.63 $\pm$ 0.01  \\
 \hline
 1 & 1.43 $\pm$  0.02   \\
 \hline
  1.25 & 1.12 $\pm$ 0.03   \\ 
 \hline
 1.5 & 0.89 $\pm$ 0.02   \\
 \hline
 1.75 & 0.60 $\pm  $ 0.02   \\
 \hline
 2 & 0.52 $\pm$ 0.03   \\
 \hline
  2.25 & 0.35 $\pm$ 0.02  \\
 \hline
 2.5 & 0.35 $\pm$ 0.02   \\ 
 \hline
\end{tabular}
\caption{Numerical $y_d$ obtained for $L=64$ and $128$ for the $q=3$ Potts model.}
\label{tableq3} 
\end{table}
In this case, we obtain again a good agreement between our numerical measurements of $y_d$ and the prediction~(\ref{ydpred}) for $a \leq 2$. For $a > 2$, the short-range disorder becomes relevant 
and indeed our numerical measurements are compatible with $\epsilon_{SR}=0.4$. 

%
\subsection{Conclusion}
In this paper we study the magnetic exponent of the long-range disordered Potts model by means of RG and Monte Carlo approaches. 
Our main result is eq.~(\ref{betanuq}) that provides the first prediction for the value of the pair correlation spin exponent of the  long-range correlated Potts model. 
This result includes as a special case the long-range correlated Ising model, see eq.~(\ref{betanuising}). Interestingly, an RG computation was carried out in \cite{dudka2016critical} for the long-range Ising. This approach allows one to derive the exponent of the square of the spin correlation function but not of the standard pair correlation exponent given here.
The prediction of eq.~(\ref{betanuq}) has been compared to previous Monte-Carlo results for the Ising model, see left part of Fig.~(\ref{fig:etaq2}), and to new Monte-Carlo results, see the right part of Fig.~(\ref{fig:etaq2}). We found a good agreement between the theory and simulations. 
Finally, we have discussed in sec.~(\ref{PottsspinMC}) some difficulties encountered in measuring the magnetic exponent. This motivated us to measure, in section~ (\ref{numericalresults}), the behavior of scaling function of the magnetization along the critical lines eq.~(\ref{duality}) for $q=1,2,3$, for small values of the disorder. This allowed the measure of the RG eigenvalues near the critical pure fixed point. The values we obtain nicely fits with the theoretical prediction of eq~(\ref{ydpred}), 
supporting then the quantum field theory set up that we implement to study these systems.
In particular, for $q=3$ we observe a clear cross-over between the long-range and short-range behavior of the disordered system. 
 \section{Acknowledgments}

We are grateful to Francesco Chippari for his early collaboration on this project. We thank Maxym Dudka for useful discussions and for pointing out relevant references.
\appendix

\section{Renormalization of the Potts spin field}
\label{RGdetail}
We apply a real space RG procedure by perturbing around a (global) conformal fixed point. By doing a  pertubative developpment, the correlation functions $\left<\left< O_1 O_2\cdots\right>\right>_{\mathcal{S}^{(n)}}$ of the action $\mathcal{S}^{(n)}$ are expressed in terms of the ones $ \mathbb{E}\left[\left<O_1 O_2\cdots\right>\right]$ associated to the action $\mathcal{S}^{\text{aux}}+\mathcal{S}^{(\alpha),\text{Potts}}$,  where the Potts and the disorder degrees of freedom are uncoupled, see eq.~(\ref{def:replicatedaction}). One has:  
\begin{equation}
\label{rencorr}
\left<\left<\left(\sum_{\gamma=1}^{n} s^{(\gamma)}_{{0}}\right)\cdots\right>\right>=\mathbb{E}\left[\left<\sum_{l_1,l_2\in \mathbb{N}} \frac{1}{l_1!l_2!}\left(\mathcal{O}^{\text{pert}}_1\right)^{l_1}\left(\mathcal{O}^{\text{pert}}_2\right)^{l_2}\left(\sum_{\gamma=1}^{n}s^{(\gamma)}_{{0}}\right)\cdots\right>\right] \; ,
\end{equation}
where:
\begin{equation}
\mathcal{O}^{\text{pert}}_1=g^{0}_{LR}\sum_{\alpha=1}^{n}\int \sigma \varepsilon^{(\alpha)},\quad \mathcal{O}^{\text{pert}}_2 =g^{0}_{SR}\sum_{\alpha\neq \beta=1}^{n}\int \varepsilon^{(\alpha)} \varepsilon^{(\beta)} \; .
\end{equation}
By relating the couplings and the operators of theories at different scale through an RG transformation, one can compute, using the expansion (\ref{rencorr}),  the renormalization (\ref{Zs}) order by order:
\begin{equation}
\label{zsseries}
Z_s=1 + \;Z^{(1)}_{s}+Z^{(2)}_{s}+\cdots \; ,
\end{equation} 
where $Z^{(n)}_{s}$ will depend in general on $\left(g^{0}_{SR}\right)^n, \left(g^{0}_{SR}\right)^{n-1}g^{0}_{LR},\cdots,\left(g^{0}_{LR}\right)^{n}$. 

The real space RG procedure requires to compute integrals over the space of many-point correlation function of the unperturbed theory $\mathcal{S}^{\text{aux}}+\mathcal{S}^{(\alpha),\text{Potts}}$. 

We recall the main results about the Potts CFT that will be used here. As usual for CFT with central charge $c\leq 1$, we parametrizes the central charge with the real number $\beta\geq 1$, $c=1-6(\beta-\beta^{-1})^2$, and the conformal dimension $\Delta_{r,s}$ with a couple of indexes $(r,s)$: 
\begin{equation}
\label{hrs}
\Delta_{r,s}=\frac{(r-1+(s-1)\beta^2)(r+1-(s+1)\beta^2)}{4\beta^2}\; .
\end{equation}
For the $q-$Potts model, $\beta$ is related to $q$ as:
\begin{equation}
\label{betaq}
\beta(q)=\sqrt{\frac{2\pi}{2\pi-\arccos{\frac{q-2}{2}}}}\; .
\end{equation} 
In this work we consider the regime $2\leq q\leq 3$, so $\sqrt{4/3}\leq \beta< \sqrt{6/5}$.

The Potts CFT spectrum has been known since longtime \cite{fsz87}. Among the Virasoro representations appearing in the Potts spectrum, one can distinguish  the ones associated to the primary fields $V_{r,s}$ that have scaling dimension $h_{r,s}=\Delta_{r,s}+\Delta_{r,-s}$ and integer spin $r s\in \mathbb{Z}$. 
The indexes $(r,s)$ take fractional values \cite{nivesvivat2024critical}, always satisfying the condition of integer spin.  It is only recently that the correlation function of these fields has been understood \cite{prs16,js18, prs19, jrs22,nivesvivat2024critical},  and the fine structure of the Virasoro representation behind have been understood \cite{js18,jrs22,NivesvivatRibault,ribault22}. 
The spin field in particular is related to the spin-less primary field $V_{0,\frac12}$,
\begin{equation}
\label{spinidentif}
s^{(\gamma)}_{0} = V^{(\gamma)}_{0,\frac12}.
\end{equation}
The scaling dimension of the spin at the pure point $h_{s}^{(P)}$ is then given by:
\begin{equation}
\label{hsP}
h_s^{(P)}=h_{0,\frac12}=\frac12 -\frac14 \beta^{-2}-\frac{3}{16}\beta^2.
\end{equation}
The identification eq.~(\ref{spinidentif}) can be shown to be equivalent to the one used in  \cite{dotsenko1995renormalisation}.

An important role is played by Potts thermal sector formed by the family of  primary denoted as $V_{\left<1,r\right>}$, with $r \in \mathbb{N}^{*}$. They are spin-less fields with scaling dimension $h_{\left<1,r\right>}=2\Delta_{1,r}$.  The correlation functions containing the  $V_{\left<1,n\right>}$ fields satisfy  $r-$order differential Fuchsian equations and admit a Coulomb gas integral representation. We use in particular this latter property to compute the correlation functions involving the energy field $\varepsilon$. The Potts energy field $\varepsilon$  is related to the field $V_{\left<1,2\right>}$,  
\begin{equation}
\label{enident}
\varepsilon^{(\gamma)}=V^{(\gamma)}_{\left<1,2\right>}\; ,
\end{equation}
and therefore:
\begin{equation}
h^{(P)}_{\varepsilon}=h_{\left<1,2\right>}=-1+\frac{3}{2}\beta^2 \; .
\end{equation}
As 
\begin{equation}
\label{esrex}
\esr= 2 -2 h^{(P)}_{\varepsilon},
\end{equation} 
 one arrives at eq.~(\ref{esrq}) and eq.~(\ref{elresra}).

The main ingredient to compute the RG flow are the fusion rules between the fields. 
We use of the following Potts CFT fusion rules :
\begin{equation}
V_{0,\frac12}\times V_{\left<1,2\right>}\to V_{0,\frac12}+V_{0,\frac32}\; .
\end{equation}
Concerning the behavior of the disorder fields $\sigma$, determined by the action $\mathcal{S}^{\text{aux}}$,  we assume that the identity field,  produced in the fusion between two $\sigma$ fields, 
\begin{equation}
\sigma\times \sigma =\text{Identity}+\cdots,
\end{equation}
 is the only one giving a contribution, in the replica $n\to 0$ limit to the renormalization equations. In \cite{Chippari23theo} we considered and discussed different type of disorder distributions where these assumptions are valid. In particular for the disorder distribution used in our simulations where the $\sigma$ are sampled from the configuration of $m-$independent  critical Ising models. More in detail, one has $\mathcal{S}^{\text{aux}}=\sum_{i=1}^{m}\mathcal{S}^{(i)}_{\text{Ising}}$ and $\sigma=\prod_{i=1}^{m}\sigma_{i}$, with the $\sigma_i$ denoting the spins of each auxiliary Ising model.

We have now all the elements to determine the RG transformations concerning the Potts spin. 
\subsection{1-loop order}
Using the identifications (\ref{spinidentif}) and (\ref{enident}), at one loop-order there are no fusions (diagrams) that can generate a single spin field: 
\begin{align}
&V^{(\alpha)}_{0,\frac12}\;\sum_{\gamma\neq \rho=1}^n V^{(\gamma)}_{0,\frac12}V^{(\rho)}_{0,\frac12}\schemestart
\arrow{-x>} 
\schemestop V^{(\alpha)}_{0,\frac12}\\
&V^{(\alpha)}_{0,\frac12}\;\sum_{\gamma=1}^n \sigma V^{(\gamma)}_{0,\frac12}\schemestart
\arrow{-x>} 
\schemestop V^{(\alpha)}_{0,\frac12},
\end{align}
which implies $Z^{(1)}=0$.

\subsection{2-loop order}
At the $2-$loop order instead, we have the following sequence of fusions:
\begin{align}
\left(g^{(0)}_{SR}\right)^2:&V^{(\alpha)}_{0,\frac12}\;\sum_{\gamma\neq \rho=1}^n V^{(\gamma)}_{\left<1,2\right>}V^{(\rho)}_{\left<1,2\right>}\sum_{\eta\neq \kappa=1}^n V^{(\kappa)}_{\left<1,2\right>}V^{(\eta)}_{\left<1,2\right>}
\schemestart
\arrow{->} 
\schemestop  V^{(\alpha)}_{0,\frac12},\quad (\alpha=\gamma=\kappa, \rho=\eta)\\
\left(g^{(0)}_{LR}\right)^2:&V^{(\alpha)}_{0,\frac12}\;\sum_{\gamma=1}^n \sigma V^{(\gamma)}_{\left<1,2\right>}\;\sum_{\rho=1}^n \sigma V^{(\rho)}_{\left<1,2\right>}\schemestart
\arrow{->} 
\schemestop V^{(\alpha)}_{0,\frac12}\quad (\alpha=\gamma=\rho).
\end{align}
Taking into account the factor of the expansions $(1/(l_1!l_2!))$ in (\ref{rencorr}) and counting the number of the diagrams and the corresponding amplitude,  we have:
\begin{equation}
\label{z2s}
Z^{(2)}_s= 2(n-1)\left(g^{(0)}_{SR}\right)^2 \mathcal{I}^{(SR)}_2+\frac{1}{2}\left(g^{(0)}_{SR}\right)^2 \mathcal{I}^{(LR)}_2,
\end{equation}
where:
\begin{align}
\label{defI3SR}
&\mathcal{I}^{SR}_2=\int_{|y|<r} d^2\;y \;\int_{|z|<r} d^2\;z \;\left< V_{0,\frac12}(0)V_{\left<1,2\right>}(y) V_{\left<1,2\right>}(z)V_{0,\frac12}(\infty)\right> \left<V_{\left<1,2\right>}(y)V_{\left<1,2\right>}(z)\right>\\
\label{defI3LR}
&\mathcal{I}^{LR}_2=\int_{|y|<r} d^2\;y \;\int_{|z|<r} d^2\;z \;\left< V_{0,\frac12}(0)V_{\left<1,2\right>}(y) V_{\left<1,2\right>}(z)V_{0,\frac12}(\infty)\right> \mathbb{E}\left[\sigma(y)\sigma(z)\right].
\end{align}
Notice that the integral $\mathcal{I}^{SR}_3$, which originates only from the short-range terms of the action,  has been fully analyzed in \cite{dotsenko1995renormalisation}. 
The four-point correlation functions can be expressed in terms of Coulomb Gas integrals with one screening, see \cite{dotsenko1995renormalisation}:
\begin{align}
\label{4ptCG}
\left< V_{0,\frac12}(0)V_{1,2}(y) V_{1,2}(z)V_{0,\frac12}(\infty)\right> &= -\frac{2\Gamma\left[-\frac{2}{3}\right]^2}{\sqrt{3} \Gamma\left[-\frac{1}{3}\right]^4}|y|^{1-\frac{1}{2\beta^2}}|y-z|^{\beta^{-2}} \\
&\; \; \qquad
\times \int d^2 u\;|u|^{-2+\frac{1}{ \beta^2}}|u-z|^{-\frac{2}{\beta^2}}|u-y|^{-\frac{2}{\beta^2}} \nonumber
\end{align}  
The integral $\mathcal{I}^{SR}_3$ has already been analyzed in \cite{dotsenko1995renormalisation}. The $\mathcal{I}^{LR}_3$ can be analyzed exactly at the same way as it is obtained by $\mathcal{I}^{SR}_3$ by replacing  the term $\left<V_{0,\frac12}(y)V_{0,\frac12}(z)\right>=|y-z|^{2-\esr}$ with $\mathbb{E}\left[\sigma(x) \sigma(0)\right]=|y-z|^{2-2\elr+\esr}$.
The analytic properties of the two integrals are therefore the same. We obtain:
\begin{align}
\mathcal{I}^{SR}_2(\esr)&= \frac{\pi^2}{8}\left(2+\xi^2\right)\;r^{2\esr} 
\label{resI2SR}\\
\label{resI2LR}
\mathcal{I}^{LR}_2(\esr,\elr)&= \left[\frac{\pi^2}{2}\elr +\frac{\pi^2}{4}\xi^2\esr\right]\;\frac{r^{2\elr}}{2\elr},   
\end{align}
with $\xi$ given in eq.~(\ref{xi}):

\subsection{3-loop order}
At the $3-$loop order instead, we have the following sequence of fusions:
\begin{align}
\left(g^{(0)}_{LR}\right)^3:&\quad V^{(\alpha)}_{0,\frac12}\;\sum_{\gamma\neq \rho=1}^n V^{(\gamma)}_{\left<1,2\right>}V^{(\rho)}_{\left<1,2\right>}\sum_{\eta\neq \kappa=1}^n V^{(\kappa)}_{\left<1,2\right>}V^{(\eta)}_{\left<1,2\right>}\;\sum_{\iota\neq \tau=1}^n V^{(\iota)}_{\left<1,2\right>}V^{(\tau)}_{\left<1,2\right>}\nonumber\\
&
\schemestart
\arrow{->} 
\schemestop  V^{(\alpha)}_{0,\frac12},\quad (\alpha=\gamma=\kappa, \rho=\iota,\eta=\tau)\\
\left(g^{(0)}_{LR}\right)^2g^{(0)}_{SR}:&\quad V^{(\alpha)}_{0,\frac12}\;\sum_{\gamma=1}^n \sigma V^{(\gamma)}_{\left<1,2\right>}\;\sum_{\rho=1}^n \sigma V^{(\rho)}_{\left<1,2\right>}\sum_{\iota\neq \tau=1}^n V^{(\iota)}_{\left<1,2\right>}V^{(\tau)}_{\left<1,2\right>}\nonumber \\
&\schemestart
\arrow{->} 
\schemestop V^{(\alpha)}_{0,\frac12}\quad (\alpha=\gamma=\iota, \rho=\tau).
\end{align}
Again, taking into account the factors of the expansions $1/(l_1!l_2!)$ in (\ref{rencorr}) and counting the number of the diagrams and  the corresponding amplitude, the total contribution to $Z^{(3)}_s$ is:
\begin{equation}
\label{z3s}
Z^{(3)}_s= 4(n-1)(n-2)\left(g^{(0)}_{SR}\right)^3 \mathcal{I}^{(SR)}_3+2(n-1)\left(g^{(0)}_{LR}\right)^2 g^{(0)}_{SR}\; \mathcal{I}^{(LR)}_3,
\end{equation}
where
\begin{align}
\label{defI4SR}
\mathcal{I}^{(SR)}_3&=\int_{|y|<r} d^2\;y \;\int_{|z|<r} d^2\;z \;\int_{|w|<r} d^2\;w\;\left< V_{\left<1,2\right>}(w)V_{\left<1,2\right>}(y)\right>\left<V_{\left<1,2\right>}(w)V_{\left<1,2\right>}(z)\right>\times\nonumber \\
&\times \left<V_{0,\frac12}(0)V_{\left<1,2\right>}(y)V_{\left<1,2\right>}(z)V_{0,\frac12}(\infty)\right> \\
\mathcal{I}^{(LR)}_3&=\int_{|y|<r} d^2\;y \;\int_{|z|<r} d^2\;z \;\int_{|w|<r} d^2\;w\;\mathbb{E}\left[ \sigma(w)\sigma(y)\right]\left<V_{\left<1,2\right>}(w)V_{\left<1,2\right>}(z)\right>\times\nonumber \\
&\times \left<V_{0,\frac12}(0)V_{\left<1,2\right>}(y)V_{\left<1,2\right>}(z)V_{0,\frac12}(\infty)\right>
\end{align}
The integral $\mathcal{I}^{(SR)}_3$ has already been considered and analyzed in \cite{dotsenko1995renormalisation}. The $\mathcal{I}^{(LR)}_3$ have the same analytic properties and can be analyzed by using the same methods.
We obtain:
\begin{align}
\label{resI3SR}
\mathcal{I}^{(SR)}_3&= \pi^3\left[3+\xi^2\right] \frac{r^{3\esr}}{3\esr} +O(\esr)\\
\label{resI3LR}
\mathcal{I}^{(LR)}_4&= \pi^3 \elr \left[\frac{2\elr +\esr \left(\xi^2+1\right) }{2 \esr\elr-\esr^2}\right]\frac{r^{2\elr+\esr}}{2\elr+\esr}+O(\esr,\elr).
\end{align}
Notice that, when $\esr=\elr$, $\mathcal{I}^{(SR)}_2=\mathcal{I}^{(LR)}_2$ and $\mathcal{I}^{(SR)}_3=\mathcal{I}^{(LR)}_3$. 

We can now collect all the results, by using (\ref{resI2SR}), (\ref{resI2LR}), (\ref{resI3SR}) and (\ref{resI3LR}) in (\ref{z2s}) and in (\ref{z3s}) to determine the $Z_s$, see (\ref{zsseries}). From the definition (\ref{gammadef}), we finally arrive at (\ref{resgamma}) by keeping all the terms till the order $g^3_{SR}, g^2_{LR}g_{SR}$ and expressing the $\gamma_s$ function in terms of the renormalized couplings.

\bibliographystyle{ieeetr}

\bibliography{biblio}

\end{document}